# Penta-Hepta Defect Motion in Hexagonal Patterns

Lev S.Tsimring
*Institute for Nonlinear Science, University of California, San Diego, CA 92093-0402*

Structure and dynamics of penta-hepta defects in hexagonal patterns is studied in the framework of coupled amplitude equations for underlying plane waves. Analytical solution for phase field of moving PHD is found in the far field, which generalizes the static solution due to Pismen and Nepomnyashchy (1993). The mobility tensor of PHD is calculated using combined analytical and numerical approach. The results for the velocity of PHD climbing in slightly non-optimal hexagonal patterns are compared with numerical simulations of amplitude equations. Interaction of penta-hepta defects in optimal hexagonal patterns is considered.

Hexagonal patterns in nonequilibrium extended systems are formed as a result of superposition of three plane waves oriented at $120^o$ with respect to each other. They appear naturally in a large variety of pattern-forming systems in fluid dynamics [1], optics [2], chemical kinetics [3], etc. The most generic defect in hexagonal pattern is so called penta-hepta defect (PHD) which is a bound state of two dislocations of opposite winding numbers on two different wave systems [1,4]. In the paper [5] we have demonstrated that the mechanism which provides binding force is the synchronization of the phase fields in the bulk of the system so the resonant condition for phases can be fulfilled everywhere except the core of the defect. If the wavenumbers of waves composing hexagonal pattern are equal to the onset value of cellular instability, PHD stays put. Meanwhile, observations of penta-hepta defects in non-Boussinesq Rayleigh-Bénard convection and in other systems demonstrate that PHDs in fact move slowly and eventually annihilate or disappear at the boundaries [6]. It is conceivable that the motion of PHDs is caused by ambient strain due to deviation of wavenumbers from onset value or by defect interaction with each other. An example of non-optimal hexagonal pattern, non-eqilateral (rhombic) hexagons, was considered recently in a number of papers [7–9]. As we shall see, even equilateral hexagonal patterns with wavenumber different from the optimal value, produce a driving Peach-Köhler force for penta-hepta defect. These non-optimal equilateral pattern were considered in [10–12].

In this Letter, the motion of PHDs in non-optimal hexagonal patterns is considered. Analysis is carried in the framework of three coupled equations for complex amplitudes of individual waves. These equations constitute the simplest heuristic model for hexagonal patterns and play essentially the same role here as complex Ginzburg-Landau equation for oscillatory media or Newell-Whitehead-Segel equation for roll patterns in isotropic media.

The order parameter of the hexagonal pattern is written in the form $a = \epsilon^{1/2}\sum_{j=1}^{3} B_j \exp(i(\mathbf{k}_j + \mathbf{K}_j)\mathbf{r}) + c.c.$ where $\epsilon$ is a small parameter characterizing the distance above onset, all three $|\mathbf{k}_i| = k_0$ , an onset wavenumber of the symmetry-breaking instability, $\mathbf{k}_1$ points at the $x$ direction (polar angle $\phi = 0$), and $\mathbf{k}_{2,3}$ point at $\phi = \pm 2\pi/3$ respectively. $\mathbf{K}_j$ are rescaled corrections to the optimal wavevectors $\mathbf{k}_j$, $\sum_{j=1}^{3}\mathbf{K}_j = 0$ from the resonance condition. Complex amplitudes $B_j$ are slow functions of $\mathbf{R} = \epsilon^{1/2}\mathbf{r}$ and $T = \epsilon t$. $B_j$ satisfy the following triplet of equations [11]:

$$\partial_T B_j = (\mu - K_j^2)B_j + B_{j-1}^* B_{j+1}^* - (|B_j|^2 + \gamma|B_{j-1}|^2 + \gamma|B_{j+1}|^2)B_i + (\mathbf{n}_j\cdot\nabla)^2 B_j + 2iK_j(\mathbf{n}_j\cdot\nabla)B_j . \quad (1)$$

Here $\mathbf{n}_j$ is the unit vector oriented along the wavevector of wave $j$, $\mathbf{n_1} = (1,0)$, $\mathbf{n_2} = (-\frac{1}{2}, \frac{\sqrt{3}}{2})$, $\mathbf{n_3} = (-\frac{1}{2}, -\frac{\sqrt{3}}{2})$, index $j$ is defined modulo 3, the coefficient of quadratic nonlinearity is rescaled to unity, $\mu$ is the rescaled supercriticality parameter, $\gamma$ is the ratio of the coefficient of cubic interaction of rolls of different orientation to the coefficient of cubic self-interaction, and $K_j = (\mathbf{n}_j\cdot\mathbf{K}_j)$. Spatial gradients are calculated with respect to slow variable $\mathbf{R}$, and asterisks denote complex conjugate. The applicability of these equations to description of real hexagonal patterns has been discussed earlier [5,11,13].

Eqs.(1) have a family of uniform stationary solutions, $B_j^0$, of which only equilateral one ($K_j = K$) can be expressed in a simple analytic form

$$B_j^0 = B_0 \equiv \frac{1 + \sqrt{1 + 4(\mu - K^2)(1+2\gamma)}}{2(1+2\gamma)}. \quad (2)$$

In a general case of different $K_j$ the amplitudes $B_j$ are different, and the solutions are known as rhombic patterns. Stability properties of nonequilateral rhombic patterns were studied recently in Ref. [7,8] .

Spontaneously formed hexagonal patterns are usually defect-ridden. Various defects have been described in the literature (see, for example, [14]). Most of them are not stable and either disappear quickly or transform into basic penta-hepta defects. As it was mentioned before, PHD is a bound state in which two of three modes have dislocations with opposite winding numbers. Without loss of generality we will consider a particular form of penta-hepta defect with positive dislocation in mode 1 and negative dislocation in mode 3. Mode 1 contains no dislocations. Corresponding solution to (1) can be written in the form $A_j = F_j(R,\phi)\, e^{i\theta_j(R,\phi)}$, where $R$ and $\phi$ are polar coordinates, $\oint_C \nabla\theta_1 d\mathbf{s} = 0$, $\oint_C \nabla\theta_{2,3} d\mathbf{s} = \pm 2\pi$, $F_{1,2}(0) = 0$, $F_{1-3}(\infty) = B_0$, and $C$ is a closed contour encircling the origin. This solution cannot be expressed in a closed analytic form. However, in the far field where all the amplitudes approach asymptotic value $B_0$, the following solution for the phase fields $\theta_j$ depending only on the polar angle $\phi$ has been found [13]:

$$\theta_1 = (1-\cos 2\phi)\frac{\sqrt{3}}{6},\ \theta_{2,3} = \pm\phi - [\frac{1}{2} - \cos(2\phi \mp \frac{2\pi}{3})]\frac{\sqrt{3}}{6} \tag{3}$$

In order to find the equations of motion for the penta-hepta defect in the non-optimal hexagonal pattern, we assume that PHD moves with a constant velocity $\mathbf{V}$. Transforming into a moving frame $\mathbf{R}' = \mathbf{R} - \mathbf{V}T$ then yields the set of stationary equations for $B_j(\mathbf{R}')$ which coincides with (1) with $\partial_T$ replaced by $-\mathbf{V}\nabla B_j$. Then we project these equations onto its two orthogonal translational modes, $\{\partial_\xi^* B_j\}$ and $\{\partial_\eta^* B_j\}$ (we choose a new coordinate frame $(\xi,\eta)$ where $\xi = X\cos\psi + Y\sin\psi$ is the coordinate along the defect motion, and $\eta = Y\sin\psi - X\cos\psi$ orthogonal to that, $\psi$ is the angle between the direction of defect motion and $X$-axis):

$$\overleftrightarrow{\mathbf{I}}\cdot\mathbf{V} \equiv \begin{pmatrix} I_{\xi\xi} & I_{\xi\eta} \\ I_{\eta\xi} & I_{\eta\eta} \end{pmatrix}\begin{pmatrix} V \\ 0 \end{pmatrix} = \begin{pmatrix} T_1 \\ T_2 \end{pmatrix}, \tag{4}$$

where $\overleftrightarrow{\mathbf{I}}$ is a mobility tensor of PHD,

$$\begin{aligned} I_{\xi\xi} &= \langle\sum_{j=1}^{3}|\partial_\xi B_j|^2\rangle,\ \ I_{\eta\eta} = \langle\sum_{j=1}^{3}|\partial_\eta B_j|^2\rangle, \\ I_{\xi\eta} &= I_{\eta\xi} = \frac{1}{2}\langle\sum_{j=1}^{3}\partial_\xi B_j\partial_\eta B_j^* + c.c.\rangle, \\ T_1 &= 2\pi[|B_2^0|^2 K_2\sin(\psi - \frac{2\pi}{3}) - |B_3^0|^2 K_3 \sin(\psi + \frac{2\pi}{3})], \\ T_2 &= 2\pi[|B_2^0|^2 K_2\cos(\psi - \frac{2\pi}{3}) - |B_3^0|^2 K_3 \cos(\psi + \frac{2\pi}{3})]. \end{aligned} \tag{5}$$

and $\langle ...\rangle = \iint ...dXdY$. All other terms from the r.h.s. of eq.(1) vanish upon integration under usual boundary conditions at infinity. Here we used the formula

$$\langle\partial_X B_j \partial_Y B_j^*\rangle - c.c. = 2\pi i\delta_j |B_j^0|^2, \tag{6}$$

where $\delta_j$ is a winding number of the dislocation at the particular mode, $\delta_1 = 0$, and $\delta_{2,3} = \pm 1$, respectively.

Equations of motion (4) indicate that the penta-hepta defect is driven by superposition of two Peach-Köhler forces corresponding to a strain at singular modes 2,3. If $K_{2,3} = 0$, defect does not move. Strain at non-singular mode 1 does not enter explicitly in the equations (4), however implicitly $K_1$ affects the amplitudes $B_j^0$.

The well known difficulty in treating equations of motions for topological defects is that integrals entering their mobilities diverge at large distances when stationary solutions are used in the integrands. In fact, with the static phase approximation solution (3) components of $\overleftrightarrow{\mathbf{I}}$ diverge logarithmically at both small and large $R$. At small $R$ the phase approximation is not valid, and the stationary solution of full amplitude equations (1) should be employed. More serious problem arizes at large $R$. Evidently, one could introduce an *ad hoc* large-scale cut-off due to finite-size effects, and therefore the mobility will be logarithmically dependent on the size of the box. This may be relevant only for small systems ($VR_{box} \ll 1$). Another possibility widely considered in the literature [15,16] for regular dislocations is to use solutions corresponding to moving defects [17]. In this case the integrals converge and therefore a finite velocity of dislocations can be found even in the large box limit ($VR_{box} \gg 1$). In the spirit of the calculation scheme [16] we assume that all $K_j$ are small, therefore velocity of PHD is also small, and the solution describing moving defect differs from the stationary one only at large distances $R \sim V^{-1} \gg 1$, where the phase approximation is well justified. Therefore, the region of integration for the components of the mobility tensor can be split into two parts; in the inner region ($R < R_0$, where $1 \ll R_0 \ll |\mathbf{V}|^{-1}$) the stationary PHD solution can be used, and in the outer region ($R > R_0$) the phase approximation can be used to simplify the task of finding the moving PHD solution. At $R \sim R_0$ static phase approximation solution (3) is applicable, so both parts should depend on $R_0$ logarithmically. After adding them together radius $R_0$ should drop out.

The details of this rather involved calculations will be published elsewhere [19]. The final expressions for the mobility components are relatively simple:

$$\begin{aligned} I_{\xi\xi} &= -B_0^2[\frac{5\pi}{2}\ln(w_1 V) - \pi\cos 2\psi\ln(w_2 V)] \\ I_{\xi\eta} &= B_0^2\pi\sin 2\psi\ln(w_3 V), \end{aligned} \tag{7}$$

where $w_1 = 1.24$, $w_2 = 1.14$, $w_3 = 2.00$ (we do not need $I_{\eta\eta}$).

Now we can substitute (5) and (7) into (4). Since we already assumed all three $K_j$ small, without further loss of accuracy we can take in (5) all three $|B_j| = B_0$ after which it is cancelled. Thus we obtain a set of two nonlinear algebraic equations for $V$ and $\psi$,

$$\begin{aligned} &\frac{5}{2}V\ln(w_1 V) + V\ln(w_2 V)\cos 2\psi = \\ &2K_2\sin(-\psi + \frac{2\pi}{3}) + 2K_3\sin(\psi + \frac{2\pi}{3}), \\ &V\ln(w_3 V)\sin 2\psi = 2K_2\cos(-\psi + \frac{2\pi}{3}) - 2K_3\cos(\psi + \frac{2\pi}{3}) \end{aligned} \tag{8}$$

In these equations there are only three $O(1)$ constants which (for small $K_j$) are functions of parameters $\mu, \gamma$ only.

Equations (4) have been written for a particular penta-hepta defect with positive dislocation in the second mode, negative dislocation in the third mod and no dislocation in the first mode. We label it as $(0,1,-1)$. Totally, there exist six distinct penta-hepta defects, $(0,1,-1)$, $(1,0,-1)$, $(1,-1,0)$, and their mirror images (conjugate defects) $(0,-1,1)$, $(-1,0,1)$, $(-1,1,0)$. The equations for $(1,0,-1)$, $(1,-1,0)$ PHDs can be obtained from Eqs.(8) by cyclic relabeling of $K_{1,2,3}$ and replacing $\psi$ by $\psi \pm \frac{2\pi}{3}$. For a conjugate PHD, the mobility tensor remains the same, but the r.h.s. of equations of motion change sign. Obviously, for conjugate defects, $V^* = V$ and $\psi^* = \psi + \pi$.

Numerical simulations of the amplitude equations (1) were performed using a split-step method. Linear parts were integrated using FFT, and nonlinear parts were calculated using Euler integrator. Typically we used $256^2$ spatial harmonics with periodic boundary conditions; physical system size was 100, and the time step was chosen 0.1. In all examples described below $\mu = 1$, $\gamma = 2$. As initial condition we take (0,1,-1) defect placed in the middle of the integration domain. To diminish an effect of periodic boundary conditions we introduce a circular ramp at $R > 0.4L$. Detailed results of numerical calculations will be given in [19]. Here we present only one pertinent example.

In the Figure 1 the magnitude and angle of the velocity vector found from (4) are plotted together with the results of direct numerical simulations versus $K_3$ for $K_1 = K_2 = 0.1$. Quantitative comparison indicates that both the direction of motion and the magnitude of velocity are in a good agreement with the theoretical analysis. We checked the importance of finite-size effects by computing velocity for a smaller system size $L = 40$. The results remained practically the same.

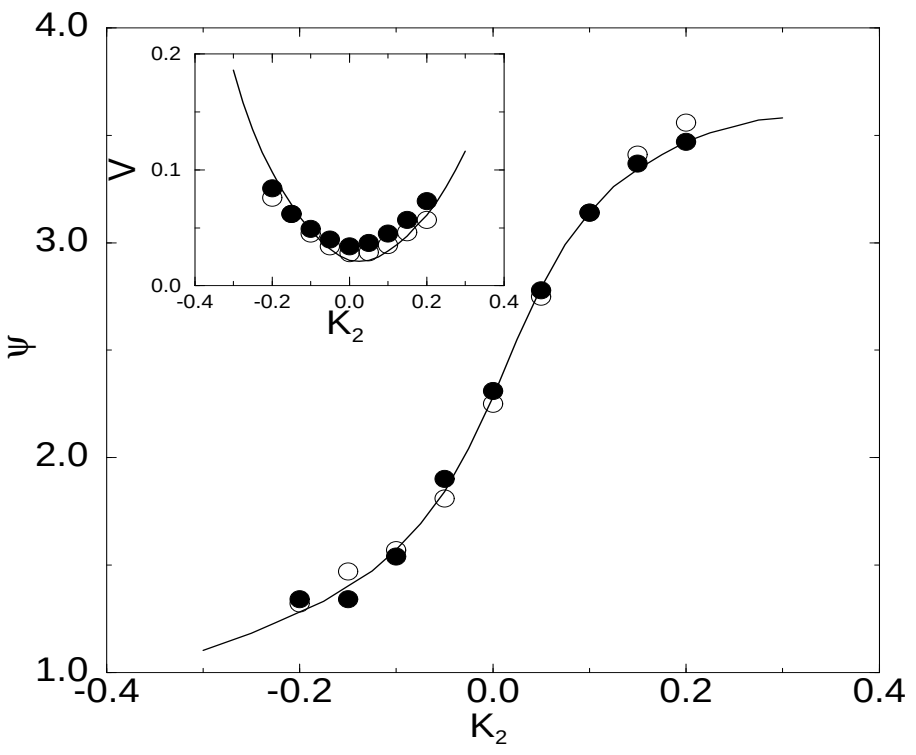

FIG. 1. Velocity vector of a PHD $(0,1,-1)$ as a function of wavenumber correction $K_2$ at $K_1 = 0.1$, $K_3 = 0.1$; main panel – angle with respect to $X$-axis, parameters $\mu = 1$, $\gamma = 2$, inset – magnitude. Solid lines – theory, Eqs.(8), open circles - numerical simulations of (1) with system size $L = 40$, solid circles - same for $L = 100$

The theory predicts (and numerical calculations confirm) that the direction of PHD motion strongly depends on the combination of the wavenumber corrections at the singular modes. On the contrary, wavenumber detuning at the non-singular mode $K_1$ only weakly affects $\mathbf{V}$ via amplitudes $B_j^0$. Notably, even if all three wavevectors are equal (but non-optimal), PHD still moves along the $X$-axis (wavevector of non-singular mode).

It is tempting to apply the equations of motion (8) directly to the interaction of two PHDs. Indeed, when two PHDs are far enough, they interact entirely through phase perturbations. Each defect distorts the phase field and therefore creates slightly non-optimal hexagonal pattern at the location of another defect. However, if we assume that the field is static, and estimate convergence velocity for a distance R, we obtain $VR \simeq 2$ which clearly is inconsistent with this assumption. Strictly speaking, phase field "remembers" whole previous path of the defects.

Here we present some of the results of numerical simulations of interacting penta-hepta defects in an ideal (all $K_j = 0$) pattern. The result of interaction (attraction or repulsion) of two PHDs, $(\delta_1^1, \delta_2^1, \delta_3^1)$ and $(\delta_1^2, \delta_2^2, \delta_3^2)$, depends on the number $N = \sum_{j=1}^{3} \delta_j^1 \delta_j^2$. It can only take values of $-2$, $-1$, 1, and 2. If $N < 0$ defects attract each other, and in all other cases they repel each other. In the Figures 2a,b two families of PHD trajectories are shown for several initial positions of defects. Figure 2a illustrates attraction of two conjugate defects, $(0,1,-1)$ and $(0,-1,1)$ (in this case $N = -2$ and defects are attracted). After collision pairs of dislocations at modes 2 and 3 annihilate, and thus perfect hexagonal pattern establishes. Meanwhile, two PHDs of the same type, $(0,1,-1)$ (here $N = 2$), repel each other (see Fig.2b). For two different PHDs, $(0,1,-1)$ and $(-1,0,1)$, $N = -1$, and defects are attracted again, however complete annihilation does not occur. Instead, conjugate dislocations at mode 3 annihilate, and remaining dislocations at modes 1 (from the first PHD) and 2 (from the second PHD) immediately form a new penta-hepta defect, $(-1,1,0)$. As this defect is alone, and the ambient strain is absent, the defect stays put.

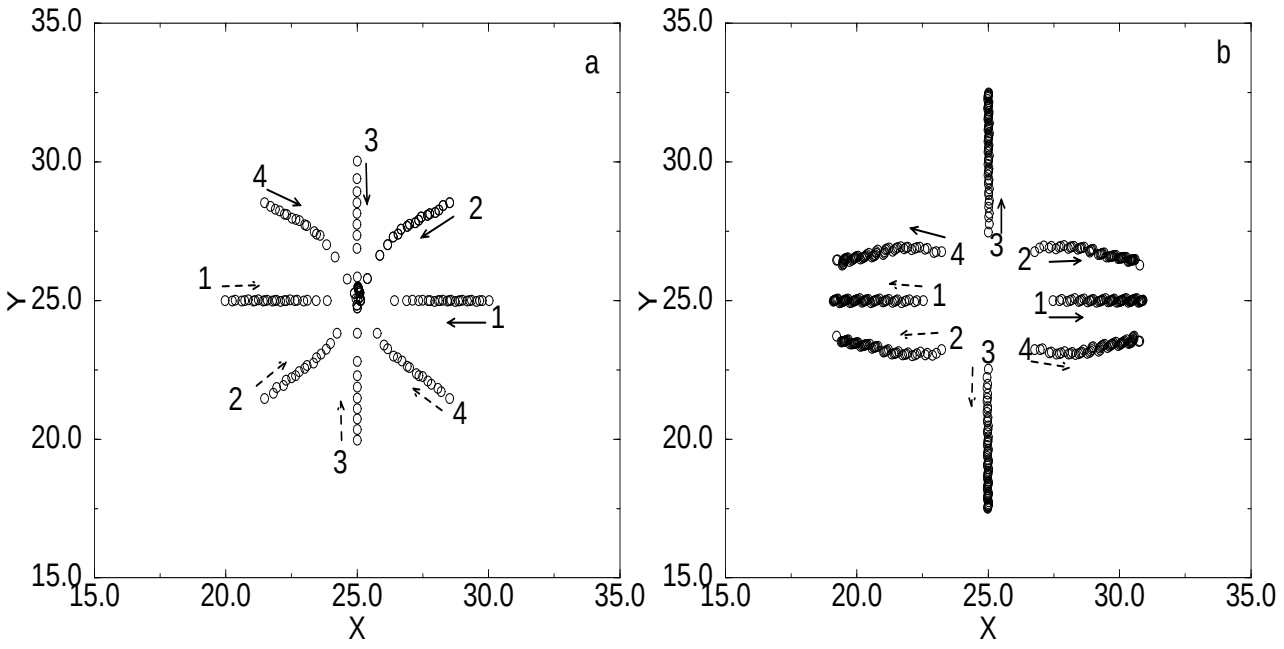

FIG. 2. Trajectories of interacting penta-hepta defects. Open circles indicate positions of the defect cores with time interval $\Delta T = 2.5$. Arrows point toward directions of motion; $a$ – $(0,1,-1)$ (solid) and $(0,-1,1)$ (dashed); $b$ – $(0,1,-1)$ and $(0,1,-1)$. For each case, four sets of initial conditions are taken (they are labeled 1-4 in the figures).

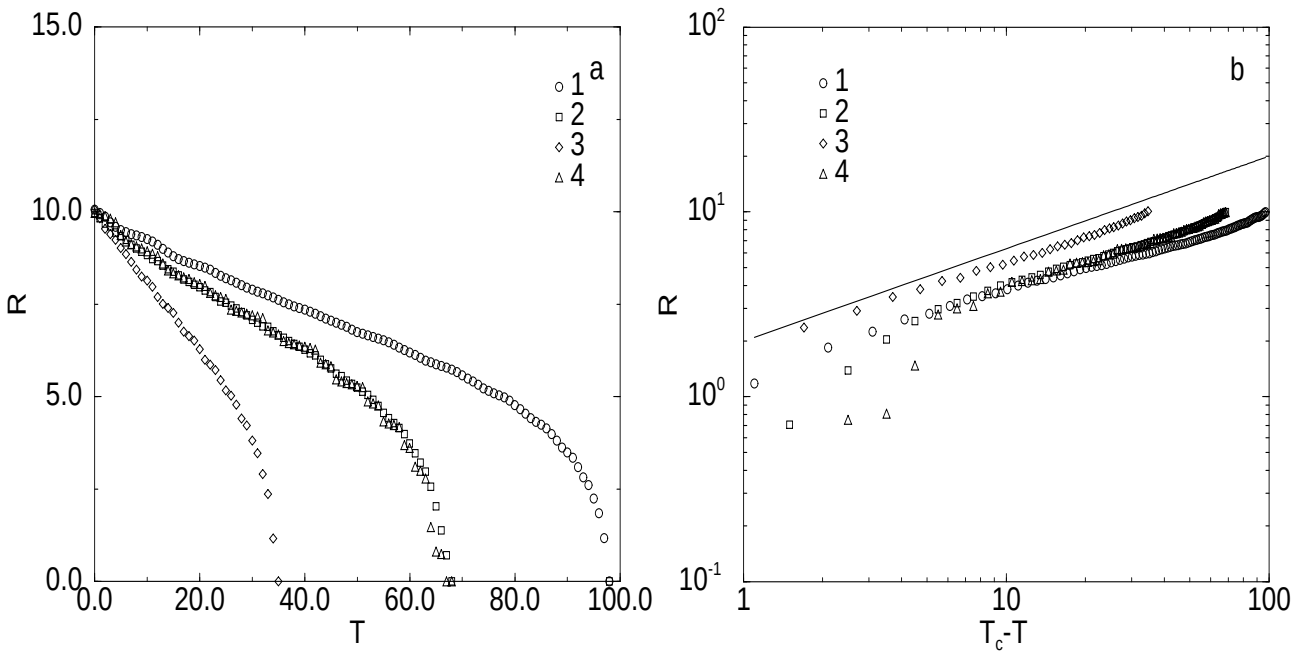

FIG. 3. Distances between interacting penta-hepta defects $(0,1,-1)$ and $(0,-1,1)$ versus time, $a$ – linear coordinates, $b$ – logarithmic coordinates. Labels 1-4 correspond to different initial positions of the defects, as shown in Fig.2a.

In the Figure 3a the distance between two cores is shown as a function of time for interacting $(0,1,-1)$ and $(0,-1,1)$ defects. The rate of convergence varies with initial positions of defects. Attraction is strongest for defects aligned along $Y$-axis, and weakest for defects aligned along $X$-axis. Figure 3b plots the same data in the logarithmic coordinates. Rather unexpectedly, one can see that over large time intervals the data is consistent with $R \propto T^{1/2}$ law, which in turn suggests $V \propto R^{-1}$ scaling. Up to logarithmic corrections this scaling is what one could expect in the static approximation discussed above. The same scaling is observed for other PHD configurations.

In conclusion, we investigated the motion of penta-hepta defects in slightly non-optimal hexagonal patterns and their interaction. PHD is stationary only in the perfect pattern with wavenumbers of all three modes equal to the onset value (in hexagonal patterns it does not correspond to the boundary of zig-zag instability). In non-optimal hexagonal patterns PHD is driven by the superposition of two Peach-Köhler forces, corresponding to two singular modes. Comparison of the theoretical predictions with numerical simulations of (1) showed good agreement in both the direction of PHD drift and the magnitude of velocity. PHDs attract or repel each other depending on the parameter $N$ introduced above. The trajectories of interacting defects may be rather complicated. Furthermore, when two attracting PHDs collide, they do not necessarily annihilate, but may give birth to another PHD with a different topological structure.

Equations (1) represent only a simplest possible model for hexagonal pattern formation. More realistic models derived from first principles (see, e.g., [9]), usually include non-variational terms. Nevertheless, we expect that major features of PHD dynamics described in this Letter will remain unchanged. We believe that detailed experiments with thermoconvection or parametric ripples similar to ones which have been performed for dislocations in roll patterns [20] could test predictions of our theory.

Author is grateful to I.Aranson, H.Levine, and M.I.Rabinovich for useful discussions. This work was supported by the U.S. Department of Energy under contract DE-FG03-90ER14138 and by the Office of Naval Research under contract N00014-D-0142 DO#15.

---

[1] E.Bodenschatz, J.R.DeBruyn, G.Ahlers, and D.S.Cannell. *Phys. Rev. Lett.*, **67**, 3078 (1991).

[2] W.J.Firth, in *SPIE Proceedings*, **2039**, "Chaos in Optics", 290 (1993).

[3] Q.Ouyang, and H.L.Swinney, *Nature*, **352** 61 (1991).

[4] S.Ciliberto, P. Coullet, J.Lega, E.Pampaloni, and C.Perez-Garcia. *Phys. Rev. Lett.* **65**, 2370 (1990).

[5] M.I.Rabinovich and L.S.Tsimring, *Phys. Rev. E*, **49**, R35 (1993).

[6] G.Ahlers, private communication.

[7] Q.Ouyang, and G.H.Gunaratne, and H.L.Swinney, *Chaos*, **3**, 707 (1993).

[8] B.A.Malomed, A.A.Nepomnyashchy, and A.E.Nuz, *Physica D*, **70**, 357 (1994).

[9] E.A.Kuznetsov, A.A.Nepomnyashchy, and L.M.Pismen, preprint, Technion, 1994.

[10] B.A.Malomed and M.I.Tribelsky, *Soviet Physics - JETP*, **65**, 305 (1987).

[11] M.M.Sushchik, L.S.Tsimring, *Physica D*, **74** 90 (1994).

[12] J.Lauzeral, S.Metens, and D.Walgraef, *Europhys. Lett.*, **24**, 707 (1993).

[13] L.M.Pismen and A.A.Nepomnyashchy, *Europhys. Lett.*, **24**, 461 (1993).

[14] J.Pantaloni and P.Cerisier, in: Cellular Structures in Instabilities, J.E.Weisfreid and S.Zaleski, eds., Springer, Berlin, 1984.

[15] E.D.Siggia and A.Zippelius, *Phys. Rev. A*, **24**, 1036 (1981); Y.Pomeau, P.Manneville, and S.Zaleski, *Phys. Rev. A*, **27**, 2710 (1983); G.Tesauro and M.C.Cross, *Phys. Rev. A*, **34**, 1363 (1986).

[16] E.Bodenschatz, W.Pesch, and L.Kramer, *Physica D*, **32**, 135 (1988).

[17] Strictly speaking, asymptotic matching procedure outlined in [18] is a better method to avoid divergencies in such problems. Unfortunately, in the present context, it does not seem feasible to implement. Note, however, that tedious asymptotic analysis of vortex motion in real Ginzburg-Landau equation [18] produced results which agreed quantitatively with simpler, albeit mathematically less rigorous approach [16] employed here.

[18] L.M.Pismen and J.D.Rodriguez, *Phys. Rev. A*, **42**, 2471 (1990).

[19] L.S.Tsimring, *Physica D*, submitted.

[20] A.Pocheau and V.Croquette, *J. Physique*, **45**, 35 (1984); R.Ribotta and A.Joets, in: Cellular Structures in Instablilities, J.E.Weisfreid and S.Zaleski eds., Springer, Berlin, 1984, p.249.